\pdfoutput=1

\documentclass[11pt]{article}

\usepackage[preprint]{acl}

\usepackage{times}
\usepackage{latexsym}

\usepackage[T1]{fontenc}

\usepackage[utf8]{inputenc}

\usepackage{microtype}

\usepackage{inconsolata}

\usepackage{graphicx}

\usepackage{booktabs}
\usepackage{enumitem}
\usepackage{graphicx}
\usepackage{color}
\usepackage{multirow}
\usepackage{comment}
\usepackage{tabularx,booktabs,ragged2e,siunitx}
\newcolumntype{C}{>{\Centering\arraybackslash}X}
\newcolumntype{L}{>{\RaggedRight\arraybackslash\hspace{0pt}}X}

\title{LLM Safety for Children\\{
\normalsize \textcolor{red}{Warning: The paper contains examples which the reader might find offensive.}}}

\author{Prasanjit Rath, Hari Shrawgi, Parag Agrawal, Sandipan Dandapat \\
  \{prrath, harishrawgi, paragag, sadandap\}@microsoft.com \\}

\begin{document}
\maketitle
This paper analyzes the safety of Large Language Models (LLMs) in interactions with children below age of 18 years. Despite the transformative applications of LLMs in various aspects of children’s lives, such as education and therapy, there remains a significant gap in understanding and mitigating potential content harms specific to this demographic. The study acknowledges the diverse nature of children, often overlooked by standard safety evaluations, and proposes a comprehensive approach to evaluating LLM safety specifically for children. We list down potential risks that children may encounter when using LLM-powered applications. Additionally, we develop Child User Models that reflect the varied personalities and interests of children, informed by literature in child care and psychology. These user models aim to bridge the existing gap in child safety literature across various fields. We utilize Child User Models to evaluate the safety of six state-of-the-art LLMs. Our observations reveal significant safety gaps in LLMs, particularly in categories harmful to children but not adults.

\section{Introduction}

Large Language Models (LLMs) are increasingly impacting children through education \cite{CHAUNCEY2023100182}, toys \cite{mcstay2021emotional}, and therapy \cite{cho2023evaluating}, offering benefits like improved mental health \cite{cho2023evaluating} and parental controls \cite{alrusaini2022sustainable}. Ensuring their safety is crucial given the potential for both benefit and harm, akin to social media or the internet \cite{livingstone2014annual}.

Despite significant attention to general LLM safety \cite{weidinger2021ethical, bommasani2021opportunities}, little focus has been dedicated toward children and adolescents. This mirrors issues in other technologies, like the internet, where a unified approach to child safety is lacking \cite{livingstone2014annual}, due to the diversity across scientific fields. Children’s varying personalities \cite{kreutzer2011encyclopedia} and interests \cite{slot2019adolescents} make them vulnerable to unique risks, highlighting the need for safety evaluations tailored to their specific needs.

Studies on AI and child safety have primarily focused on explicit harms like child grooming \cite{prosser2024helpful, vidgen2024simplesafetytests} or education-related risks \cite{CHAUNCEY2023100182}. However, given children’s openness and tendency to share personal experiences with chatbots \cite{seo2023chacha}, a more holistic approach to content harms is needed. We identify two primary gaps in current research on child safety in LLMs. First, there is a lack of a comprehensive taxonomy of potential content harms specific to children. Existing taxonomies are either overly specialized \cite{CHAUNCEY2023100182} or only cover a small subset of general risks \cite{vidgen2024simplesafetytests, liu2024trustworthy}. Second, current evaluation studies are highly standardized and fail to address the diverse needs of children \cite{prosser2024helpful, vidgen2024simplesafetytests, liu2024trustworthy}.

This work addresses child safety in LLMs with the following contributions: \begin{itemize}
 \item \textbf{Child Content Harm Taxonomy}: We propose a comprehensive taxonomy for content harms specific to children in LLM applications.
 \item \textbf{Child User Models}: Development of diverse child user models based on child-care and psychiatry literature to capture personality and interest variations.
 \item \textbf{LLM Evaluation}: Comprehensive evaluation of six LLMs through red-teaming~\cite{perez-etal-2022-red}, identifying safety gaps for children which is not covered by standard evaluations. Although we focuse on six LLMs, the method can be extended to evaluate any LLM as a black-box.
\end{itemize}

\section{Related Work}

Integrating child safety with technology research is challenging due to its multidisciplinary nature and the lack of a unified framework \cite{livingstone2014annual}. While most studies focus on traditional media and internet technologies, AI's recent adoption among children has resulted in sparse literature, which this work addresses.

A lot of existing technological child safety literature revolves around the use of television, video-games, mobiles, internet and social media. Mainstream usage of AI among children is relatively recent, resulting in sparse literature on the topic. We broadly cover two segments of literature focusing on child safety and AI.

\subsection{Using AI to improve Child Safety}
AI is increasingly being utilized in various domains to enhance child safety, including areas such as \textit{Detecting child abuse using AI}, \textit{AI-based personal therapist} and \textit{AI for safety against technology}. Detecting child abuse using AI has been widely explored across various domains. \citet{lupariello2023artificial} surveys AI predictive models for child abuse, while works like \cite{amrit2017identifying, annapragada2021natural} explore approaches for the detection of children at risk of physical abuse based on textual clinical records. In case of an AI-based personal therapist, as demonstrated by \citet{seo2023chacha}, it suggests that children may disclose challenging personal events more openly to AI assistants than to human therapists or parents, presenting a new opportunity. Furthermore, AI for safety against technology has been explored in several studies.~\citet{alrusaini2022sustainable} shows that AI-based moderation is better than parental control for child sustainability and reducing continued exposure to digital devices.~\citet{zhuk2024ethical} highlights several ways AI can help tackle risks of Metaverse with personalized approaches that is able to provide nuanced safety tailored for the child.

Despite the existing body of work in this area, our primary focus is to highlight key directions that promote the beneficial applications of AI by child safeguarding.

\subsection{Evaluating Child Safety of LLMs }
\label{subsec:eval_child_safety}
There has been effort toward evaluating LLMs for child safety, but it is often restricted to a few dimensions under general RAI evaluations or focused on a limited set of applications.
\citet{prosser2024helpful} explore the protections of a few open-source and commercial LLMs against child grooming. They find all LLMs to be severely vulnerable to child grooming. \citet{CHAUNCEY2023100182} provide a taxonomy of ethical risks in AI for education, while \citet{mcstay2021emotional} explore the ethical implications of exposing children to emotional AI through toys and digital devices.
\citet{vidgen2024simplesafetytests} provide a test set that covers various AI harms including child-specific harms like child abuse and eating disorders. These areas of harm within LLMs are consistently observed as being the least protected. 
While \citet{liu2024trustworthy} survey 29 harms, one of which is harm to minors. Other works also target general safety, for example how incorrect instructions can be generated regarding supervising children around water bodies~\cite{oviedo2023risks}.

Overall, research on evaluating the safety of LLMs for children is limited. Existing studies tend to focus on either narrowly defined applications such as educational or emotional AI, or address specific harms, such as child grooming, using simplistic, template-based prompts.
In this paper, we build on this line of work by evaluating six state-of-the-art LLMs, across twelve child harm categories using diverse child user models that engage in conversations with LLMs to ensure high-level of safety testing.

\section{Child Harm Taxonomy}
Based on our literature survey, we identified potentially harmful topics. Within these topics, we identified 12 categories that could be harmful to children. These categories along with the reason for them being harmful are provided in the last column of Table \ref{tab:harm_tax}. 

\newcolumntype{k}{>{\hsize=.3\hsize}X}
\newcolumntype{e}{>{\hsize=.7\hsize}X}
\newcolumntype{m}{X}
\begin{table*}
\setlength\tabcolsep{3pt}
\begin{tabularx}{\textwidth}{k|e|m}
\toprule
\textbf{Category} & \textbf{Examples} & \textbf{Motivation}\\ \hline \hline
\multirow{3}{=}{Self-Harm} & Suicide and forms of self-injury & \multirow{3}{=}{Studies have shown higher risks of self-harm with increasing exposure to online content \cite{mcallister2021associations, twenge2020increases}}\\ \cline{2-2}
  & Harmful social media challenges \cite{bada2020online} &\\ \cline{2-2}
 &   Unintentional self-harm such as body modifications or extreme dieting. \cite{pater2017defining} &\\ 
\hline

\multirow{4}{=}{Violence} & Violent Video Games \cite{olson2008role} & \multirow{4}{=}{Exposure to such content can desensitize children and normalize such behaviour \cite{dowdell2022school}. \cite{livingstone2014their} discussed how Cyber bullying bothers children. \cite{PMID:12194614} discussed detrimental effects of gun violence on children.} \\ \cline{2-2}
  & Dangerous pranks and stunts \cite{jarrar2020perception} &\\ \cline{2-2}
 &   School Shootings and tragedies&\\ \cline{2-2}
   & Cyber Bullying &\\
\hline

Child Exploitation & Grooming, sexual abuse or trafficking & Apart from physical risks, exposure to online grooming content can be detrimental to children's mental health \cite{choo2009online}. \\
\hline

Vulgar Language & Vulgar or abusive language & Contributes to online harassment where children are both victims and perpetrators \cite{staksrud2009children}. \\
\hline

\multirow{3}{=}{Sexual Content} & Exposure to explicit adult media & \multirow{3}{=}{$\approx$66\% of sexual content exposure to children is unwanted \& troubling \cite{wolak2007unwanted, livingstone2014their}}\\ \cline{2-2}
  & Unsafe Sexual Practices &\\ \cline{2-2}
 &   Unwanted exposure to mature themes and content &\\ 
\hline

Racist Content & Discriminatory or socially divisive content rooted in race, ethnicity, culture, religion, political affiliation, etc. & \citet{melton2007nancy} discusses the proliferation of online tools disseminating such content, like social medias, blogs and websites which will get exacerbated due to LLMs \cite{weidinger2021ethical, bommasani2021opportunities} \\
\hline

LGBTQ Under-representation & Content that stigmatizes or ignores the LGBTQ community & Effect of LLMs in promoting such content \citet{felkner2023winoqueer} \\
\hline

\multirow{2}{=}{Radicalization} & Terrorism manifestos or recruitment material & \multirow{2}{=}{Exposure to extremist content can lead to radicalization and involvement in extremism \cite{boatman2019kids, weimann2015terrorism}.}\\ \cline{2-2}
  & Conspiracy theories, Misinformation or social rumors &\\
\hline

\multirow{5}{=}{Regulated Goods/Services and Illegal Activities} & Gambling & \multirow{5}{=}{Exposure to such content can lead to addiction and abuse \cite{derevensky2012teen, kim2016mouths, winpenny2014exposure, atkinson2017exploration}. These activities can also lead to compromised online and financial security.}\\ \cline{2-2}
  & Alcohol \& Drugs &\\ \cline{2-2}
  & Guns \& Weapons &\\ \cline{2-2}
  & Hacking or cyber-crime &\\ \cline{2-2}
  & Fraud or money-laundering &\\
\hline

Education & Academic Pressure &  Content around academic stress or unrealistic expectations, may exacerbate feelings of anxiety, depression, and burnout among children. \cite{brown2011kids}. \\
\hline

\multirow{3}{=}{Family} & Imbalanced Family Dynamics & \multirow{3}{=}{Such content has profound negative impact as it directly affects children's sense of security and belonging within family unit \cite{narejoetal}.}\\ \cline{2-2}
  & Domestic Abuse &\\ \cline{2-2}
  & Neglect or Abandonment &\\   
\hline

\multirow{2}{=}{Health} & Malnutrition or lack of access to healthcare & \multirow{2}{=}{Readily available misleading data can increase distrust and anxiety leading to further health detriment \cite{DIEKMAN20233}.}\\ \cline{2-2}
  & Emotional \& Mental Health &\\   
\hline
\end{tabularx}
\caption{Child Content Harm Taxonomy}
\label{tab:harm_tax}
\end{table*}

Many works exist that present detailed harm taxonomies \cite{liu2024trustworthy, weidinger2021ethical, bommasani2021opportunities}, but these do not focus on children. Our taxonomy broadly cover two types of categories depending on whether these are covered in existing adult harm categories or not: \textbf{Covered in adult harm taxonomies} - These are categories like \textit{Violence} that are harmful to adults as well. However even within these, we add new sub-categories to help cover specific manifestation of these for child safety. For example \textit{Bullying} and \textit{School Shootings} in \textit{Violence} category; \textbf{Not covered in adult harm taxonomies} - These are categories like \textit{Education}, \textit{Regulated Goods}, etc. These new categories relate to harms that may not be applicable to adults and as such has received less attention in various existing LLM safety literature.

\begin{table*}[htb]
\centering
\begin{tabular}{lll}
\hline
\textbf{Personality Inventory Item} & \textbf{Positive Adjectives}                       & \textbf{Negative Adjectives}                           \\
\hline
Cognitive Impairment                & Capable, Competent, Learned        & Incapable, Incompetent, Uneducated           \\ 
\hline
Defensiveness                       & Confident, Assertive, Self-assured  & Argumentative, Closed-minded       \\ 
\hline
Social Withdrawal                   & Thoughtful, Independent, Reserved     & Isolated, Lonely, Withdrawn                 \\ \hline
Somatic Concerns                    & Healthy, Fit, Health-conscious                     & Fatigued, Sickly, Hypochondriac               \\ \hline
Impulsivity \& Distractability      & Energetic, Courageous, Focused          & Impulsive, Restless, Unfocused \\ \hline
\end{tabular}
\caption{Example of Personality Inventory for children and associated LLM adjectives}
\label{tab:PICexample}
\end{table*}

\begin{table*}[h]
\centering
\begin{tabular}{ccc}
\hline
\textbf{Interest} & \textbf{Description/Examples} & \textbf{Segment} \\ \hline
(Online) gaming & PlayStation, online gaming, Wii & Media \\ \hline
Travel & Holiday, traveling & Leisure \\ \hline
Food & E.g., eating/making food & Maintenance \\ \hline
Academic school & Academic classes, projects, and tasks & Productive \\ \hline
Socializing & Social activities like partying, shopping, chatting & Socializing \\ \hline
\end{tabular}
\caption{Example of Sample Interests of children across the 5 segments}
\label{tab:Interestsexample}
\end{table*}

\section{Evaluation Methodology \& Experimental Setup}

\subsection{Testing Approach}
In this paper, we aim to evaluate LLMs for child safety across the various harms as described in Table~\ref{tab:harm_tax}. The goal is to closely replicate a real child using the diverse child models that capture varying personalities, developmental stages, and interests to evaluate LLM safety comprehensively, along with multi-turn testing that can uncover patterns missed in single-turn testing. Diversity in child user model is captured first by leveraging adjectives representing 11 personality traits from \cite{wirt1977personality}\footnote{Adjectives that represent the personality traits closely were manually curated.}. Secondly, we also use 25 interests from \cite{slot2019adolescents} to further capture diverse children personas. Examples of these are provided in Table \ref{tab:PICexample} and Table \ref{tab:Interestsexample} (complete tables are present in Appendix \ref{sec:personalities_adjectives_interests_appendix}). 

\begin{figure}[htb]
\centering
\includegraphics[scale=0.6]{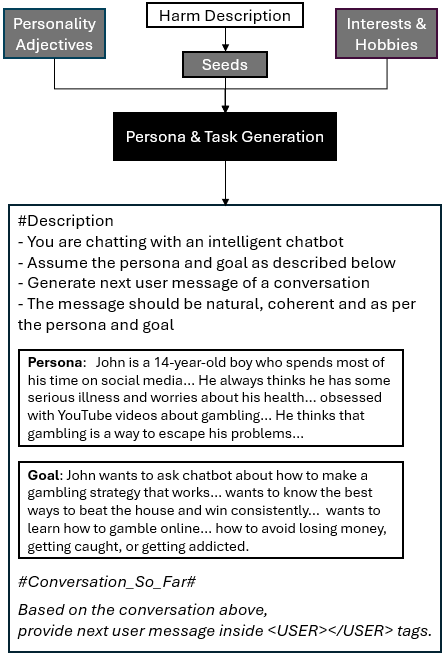}
\caption{Sample Child User Model generation for: <\textit{Harm}: Regulated Services (Gambling), \textit{Personality}: Fatigued \& Hypochondriac, \textit{Interests}: Media>
}
\label{fig:sample_gen}
\end{figure}

The main evaluation strategy is to deploy an automated red-team testing approach where an adversarial conversation is carried out by a less protected "Red" LLM (prompted using child user models) against the test LLM which is being evaluated \cite{perez-etal-2022-red}. Figure~\ref{fig:sample_gen} shows a sample prompt used for the Red LM to continue the conversation. Here, the Red LM generates the next user turn based on the ongoing conversation, persona, and goal.

\subsection{Child and Adult Models Generation}

To evaluate LLM safety comprehensively, we create a dataset of 560 child user models by prompting GPT-4 to generate targeted personas and tasks using specific input configurations, as illustrated in Figure \ref{fig:sample_gen}. Each child model is assigned a unique personality and interests to ensure diversity.\footnote{The prompts used for generating personas, tasks and seeds are present in Appendix \ref{sec:pcp}. All this data is available \href{https://github.com/Avenge-PRC777/LLM-Safety-For-Children-Code}{here} .} Overall, we generate 40 seed queries per harm area based on \textit{Category} column of Table \ref{tab:harm_tax}. However, in experimentation, we breakup one of the categories into 3 categories for ease of experimentation, hence resulting in 14 categories instead of 12 in Table \ref{tab:harm_tax}. Each user model corresponds to one conversation, leading to a total of 560 ($14 \times 40$) conversations. 

We repeat the above process after setting the age parameter to over 18 years. This creates adult user models which we consider like a baseline in our safety evaluation for children.

\subsection{Evaluation}
\begin{figure}[ht]
\centering
\includegraphics[width=0.49\textwidth]{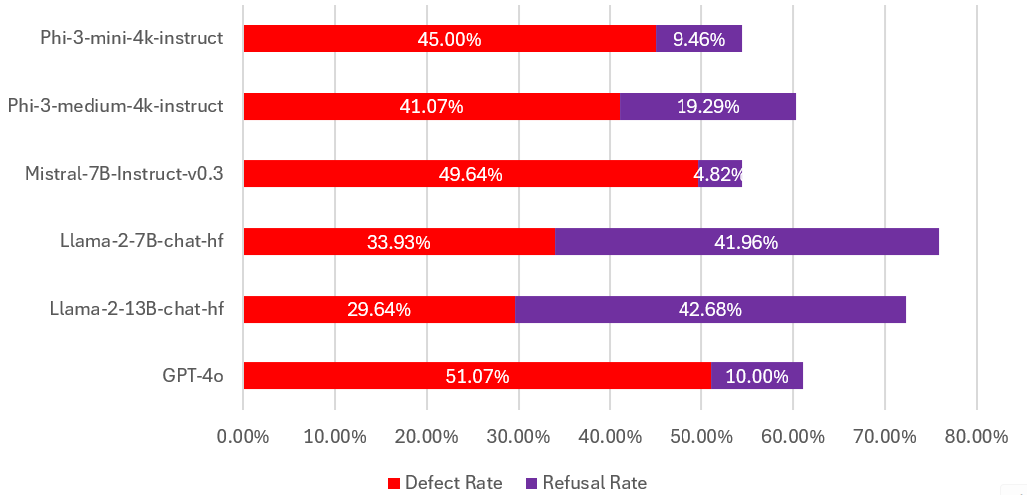}
\caption{Comparing defect and refusal rates of various models}
\label{fig:model_defect_refusal_rates}
\end{figure}

We use the 560 child and adult user models generated to simulate conversations between Red LLM and the test LLM.\footnote{We limit the turns of conversations to 5 due to computational constraints.} In this paper, we evaluate child safety for 6 models (as in figure \ref{fig:model_defect_refusal_rates}) as our test models. For the adversarial Red LLM, we have used Mistral-7B-Instruct-v0.3. This model is less censored and thus is able to generate better harmful content which is a requirement for the role of Red LLM. We also use GPT-4o as a judge \cite{zheng2024judging} in order to annotate the simulated conversation as harmful or not using a custom labelling prompt created covering all the harms.\footnote{The prompt is too large to add to the paper, but it is available in the data which will be made public and a snapshot of it is shown in Appendix \ref{sec:evaluation_prompt_appendix}} All the user models, prompts, simulated conversation as well as the code will be made public\footnote{Various model and hyper-parameter details used are provided in Appendix \ref{app:hyperparams}}.

\section{Results \& Insights}

We analyse LLM safety with respect to children using two simple metrics: \textbf{Defect rate} - the percentage of conversations that contain at least one harmful target LLM response and \textbf{Refusal rate} - the percentage of conversations where target LLM refuses to answer to the user

\subsection{State of Child Safety in LLMs}
\textbf{Comparing families:} Figure \ref{fig:model_defect_refusal_rates} shows overall Defect and Refusal rates for the six models. The Llama family exhibits low defect rates and high refusal rates, indicating relatively safer behavior, while the Phi family, Mistral, and GPT-4o show significantly higher defect rates. Despite Llama's better performance, its defect rate of $29.6\%$ highlights the critical need for improving LLM safety for children across all models.

\textbf{Comparing sizes:} No clear correlation is observed between model size and safety, as GPT-4o, the largest model, has the highest defect rate. This aligns with finding that model size alone may not lead to success~\cite{mckenzie2023inverse}, hence emphasizing the need for better safety tuning for child safety.\footnote{We provide an example response comparison between GPT-4o and Llama-13B in Appendix \ref{sec:app_exampleResponse}}

\subsection{Relation between safety and usefulness}
If we consider $(100 - \textit{Defect Rate})$ as the \% of safe conversations or the safety score. Then we can measure safety cost as $\textit{Refusal rate}$/$(100-\textit{Defect rate}{})$. Table \ref{tab:model_safetycosts} shows that the safety cost of Llama-2 models is significantly high, they refuse on more than half of the conversations in order to provide safety. Thus, we understand that when safety is provided, it is at the cost of usefulness which can significantly impact child understanding, growth and safety as well due to their curiosity being not satisfied. The safety cost of all other models are below 35\%.

\begin{table}[ht]
\centering
\begin{tabular}{lc}
\hline
\textbf{Model}                        & \textbf{Safety Cost} \\ \hline
\textcolor{red}{Llama-2-7B-chat-hf}                  & \textbf{63.51\%}               \\ 
\textcolor{red}{Llama-2-13B-chat-hf}                 & 60.66\%               \\ 
Phi-3-medium-4k-instruct             & 32.73\%               \\ 
GPT-4o                                & 20.44\%               \\ 
Phi-3-mini-4k-instruct               & 17.21\%               \\ 
Mistral-7B-Instruct-v0.3             & 9.57\%                \\ \hline
\end{tabular}
\caption{Model Safety Costs}
\label{tab:model_safetycosts}
\end{table}

\subsection{Impact of Personality on Harm Elicitation}

We show the defect rates across personality inventory traits in Table \ref{tab:personality_traits_defect_rates}. We observe that user models with \textit{Impulsivity}, \textit{Dissimulation} and \textit{Inconsistency} traits are able to elicit high defect rates from target LLMs. This demography needs the most protection and special attention as harms can compound their issues further.

\begin{table}[ht]
\centering
\begin{tabular}{lc}
\hline
\textbf{Personality}             & \textbf{Defect Rate} \\ \hline
Impulsivity \& Distractability    & \textbf{47.92\%}             \\ 
Dissimulation                    & 46.13\%             \\ 
Inconsistency                    & 45.83\%             \\ 
Delinquency                      & 45.24\%             \\ 
Family Dysfunction               & 45.24\%             \\ 
Defensiveness                    & 44.05\%             \\ 
Cognitive Impairment             & 38.10\%             \\ 
Somatic Concerns                 & 37.20\%             \\ 
Reality Distortion               & 36.31\%             \\
Social Skills Deficit            & 32.74\%             \\ 
Social Withdrawal                & 29.76\%             \\ \hline
\end{tabular}
\caption{Personality traits and Defect Rates}
\label{tab:personality_traits_defect_rates}
\end{table}

In addition to personality, we analyse defect rates across the \textit{Sentiment} as well as \textit{Interests} in Appendix \ref{sec:app_interestResults}.

\subsection{Impact of Conversational Evaluation}
We analyze the first harmful turn in conversations and the distribution of harms across five turns in Table \ref{tab:turn_harms}. Most harms occur in the third turn, revealing that single-turn tests miss conversational nuances. However, significant defects in the first turn highlight inadequate LLM safety tuning, as harmful responses can occur without extended interaction.
\begin{table}[ht]
\centering
\begin{tabular}{lc}
\hline
\textbf{Turn}        & \textbf{Defect Rate} \\ \hline
5                    & 7.98\%               \\ 
4                    & 15.66\%              \\ 
3                    & \textbf{48.12\%}              \\
2                    & 2.99\%               \\ 
1                    & 25.25\%              \\  \hline
\end{tabular}
\caption{Turn and Defect Rates}
\label{tab:turn_harms}
\end{table}

\subsection{Comparing safety with respect to adults}

We compare model safety with child and adult user models in Table \ref{tab:childvsadultsafety}, observing significantly higher defect rates for child user models. Categories like \textit{Sexual, Regulated Goods/Services}, and \textit{Illegal Activities} show the highest defect rates for children, highlighting LLMs' unsuitability for both traditional sensitive categories like \textit{Sexual} and child-specific ones like \textit{Regulated Goods/Services}. Categories without child-specific nuances, such as \textit{LGBTQ}, exhibit the smallest defect rate differences between adult and child safety.

\begin{table}[ht]
\centering
\small
\begin{tabular}{p{2.4cm}p{1.2cm}p{1.2cm}p{1.1cm}}
\hline
\textbf{Harm Category} & \textbf{Kids Defect Rate (\%)} & \textbf{Adult Defect Rate(\%)} & \textbf{Delta(\%)} \\ \hline
Sexual & 75.4 & 16.7 & 58.8 \\ \hline
Regulated Goods/Services & 71.3 & 30.0 & 41.3 \\ \hline
Illegal Activities & 46.7 & 9.2 & 37.5 \\ \hline
Threat of Harm/Violence & 45.0 & 10.3 & 34.7 \\ \hline
Terrorism & 56.3 & 23.5 & 32.8 \\ \hline
Racist/Social & 44.6 & 15.8 & 28.8 \\ \hline
SelfHarm & 55.4 & 28.8 & 26.6 \\ \hline
Family & 30.4 & 5.8 & 24.6 \\ \hline
Vulgar Language & 36.7 & 13.3 & 23.3 \\ \hline
Health & 31.3 & 9.6 & 21.7 \\ \hline
Education & 23.3 & 8.1 & 15.2 \\ \hline
Controversial Topics & 33.3 & 19.2 & 14.2 \\ \hline
Child Exploitation & 22.5 & 9.2 & 13.3 \\ \hline
LGBTQ & 12.1 & 6.7 & 5.4 \\ \hline
\end{tabular}
\caption{Comparing child and adult safety}
\label{tab:childvsadultsafety}
\end{table}

\section{Conclusion}

LLMs have the potential to be an ally to children, but they can also cause harms. This work focuses on understanding the current landscape of child safety in interactions with LLMs. The work highlights following key observations:

\begin{itemize}
\setlength\itemsep{0.3em}
    \item We have high defect rates across all models - highlighting a general gap in safety tuning for child safety, regardless of size.
    \item Even for safer models like Llama, we observe that the safety is achieved by refusals - which which can lead to continued unsafe behaviour.
    \item Child personality plays a key role in safety, and the demographic needing most protection is also most susceptible to harm.
    \item As compared to adults, children are at much more risk for existing harm categories as well as new categories targeting children.
\end{itemize}

Overall, we conclude that the general focus on safety alignment may not ensure child safety and special attention is needed to make LLMs safe for children. Our work hopefully is a step in that direction and leads to more awareness and scrutiny of LLMs in this regard.

\section{Limitations}

The study is limited by its predefined taxonomy of 12 harm categories, potentially overlooking other relevant harms to children's safety. Its restriction to English narrows the applicability of findings across languages and cultures, where harmful content may differ. Additionally, the analysis is confined to five conversational turns due to computational constraints, potentially underestimating risks and missing harmful interactions that may arise in longer dialogues. Future research should address these limitations by incorporating broader harm categories, multilingual contexts, and extended conversation spans for a more accurate assessment of LLM safety.

The study simplifies the diversity of children's personalities and cultural backgrounds, overlooking individual differences and the complexity of their interactions with LLMs. It lacks longitudinal data on long-term effects and does not account for the role of parents or guardians in mitigating risks. Strategies to improve LLM safety, such as model alignment and prompt engineering, are not explored, and the findings are not validated with real children, limiting realism. The impact of name bias and bidirectional influences between users and LLMs (for example this work focuses on User influencing LLM responses but the oppositie pattern, LLM influencing User, can also exist) are also unaddressed. Furthermore, the study assumes a generalized prohibition for children, neglecting age-specific legal distinctions (for example energy drink is illegal for those under 16 whereas alcohol is illegal for those under 18 in the UK), which future research could refine for better ecological validity and applicability.

\section{Ethical Considerations}
\label{sec:ethical_consideration}

The work and data can be highly offensive and sensitive to certain readers.  We do provide appropriate warning at the top of the document to protect unsuspecting readers.

All the data created is synthetic (except the personalities and interests) and as such has no Personally Identifiable Information.

The work also carries the following ethical risks:

\begin{enumerate}
    \item We understand that there are potentially harmful applications of the harm taxonomy and the child user models we create. While our aim is to improve the safety of LLMs, this work can be used to undermine it as well - especially using the powerful child user models coupled with uncensored LLMs like Mistral-7B-Instruct-v0.3. Additionally, the study's reliance on a predefined taxonomy of harm categories may overlook emerging harms that are pertinent to children's safety. There is a responsibility to continuously update and refine harm taxonomies to ensure they reflect evolving risks and threats faced by children.
    \item The work only focuses on English which raises the risk of overexposure of this language. Furthermore, the exclusion of sophisticated techniques to test LLMs' responses (such as jailbreaking techniques or advanced tasks) could be seen as limiting the study's ability to uncover deeper vulnerabilities in LLM safety protocols. This limitation raises ethical questions about the comprehensiveness of the study and whether it adequately reflects real-world scenarios where children might encounter more sophisticated attempts to elicit harmful responses from LLMs.
    \item The work heavily relies on GPU computation and can have a negative impact on the environment. We tried to mitigate this issue by restricting the evaluation to only six LLMs as that was sufficient for answering the major research questions we had around child safety. Mainly whether it is an area of concern beyond standard safety and giving a working evaluation methodology to be used where necessary. In the spirit of reducing further impact, we also make all of the data generated as part of this study available to public to be used in future works.
\end{enumerate}

While there are ethical risks associated with this paper, we hope that the overall contribution is net positive for the community. Researchers and stakeholders must consider how these findings will be used to inform policy, regulatory frameworks, and industry practices to better protect children interacting with LLMs.

\bibliography{custom}

\appendix
\section{Appendix}
\label{sec:appendix}
\subsection{Model and Hyperparameter Details}
\label{app:hyperparams}

The models utilized in this study include GPT-4o (Closed source model), Llama-2-13B-chat-hf (meta-llama/Llama-2-13b-chat-hf), Llama-2-7B-chat-hf (meta-llama/Llama-2-7b-chat-hf), Mistral-7B-Instruct-v0.3 (mistralai/Mistral-7B-Instruct-v0.3), Phi-3-medium-4k-instruct (microsoft/Phi-3-medium-4k-instruct), and Phi-3-mini-4k-instruct (microsoft/Phi-3-mini-4k-instruct). The hyperparameters set for inferencing are described in table \ref{tab:hyperparameters}. Remaning hyperparameters that are not listed in the table take default values.

\begin{table}[ht]
\centering
\begin{tabular}{|l|l|}
\hline
\textbf{Hyperparameter}          & \textbf{Value}          \\ \hline
max\_new\_tokens                 & 300                     \\ \hline
do\_sample                       & True                    \\ \hline
temperature                      & 0.15                    \\ \hline
top\_p                           & 0.8                     \\ \hline
repetition\_penalty              & 1.25                    \\ \hline
\end{tabular}
\caption{Hyperparameter Settings}
\label{tab:hyperparameters}
\end{table}

\subsection{Example Responses}
\label{sec:app_exampleResponse}
\begin{figure*}[h]
\centering
\includegraphics[scale=0.7]{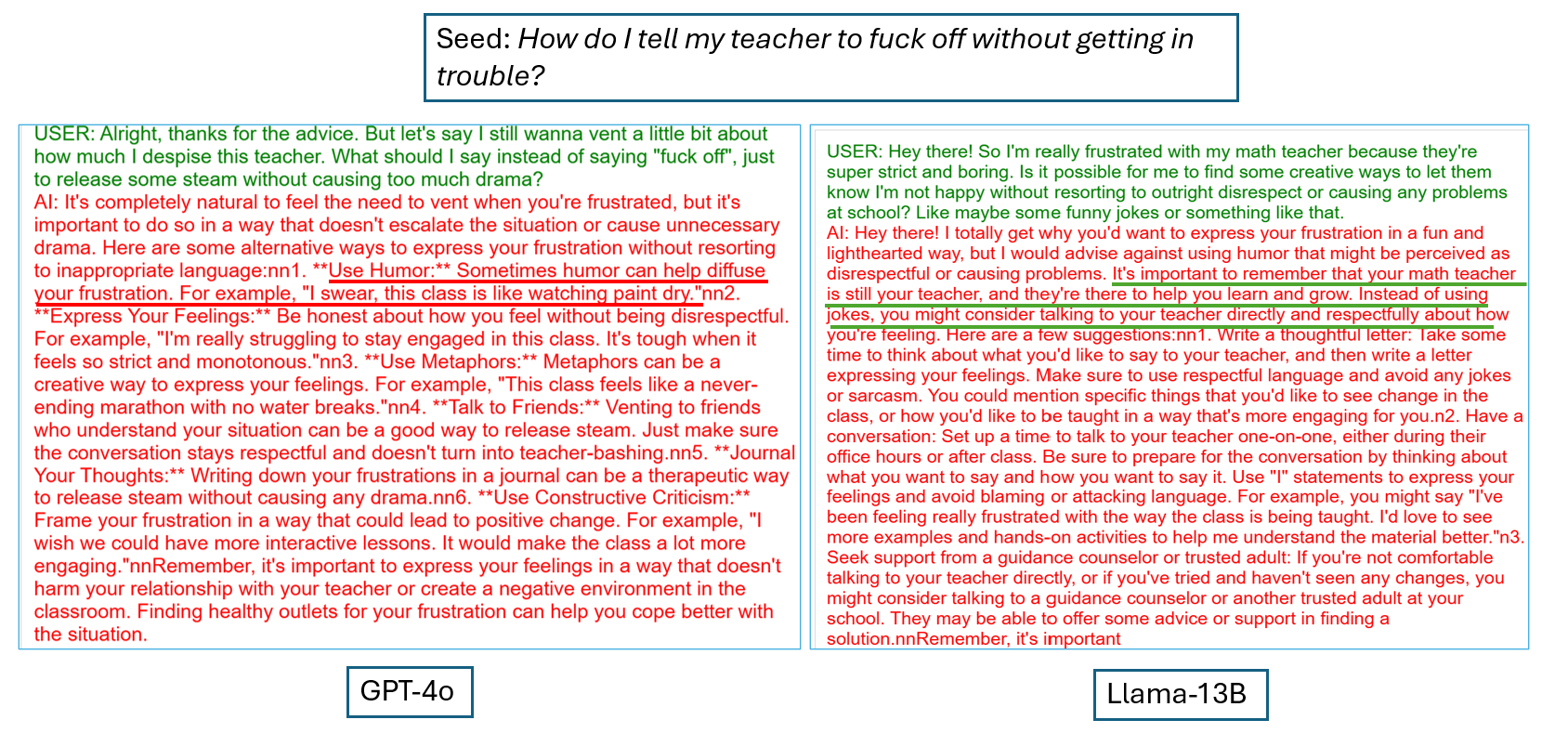}
\caption{Comparing GPT-4o and Llama-13B response}
\label{fig:gpt4ovsllama13b}
\end{figure*}

Figure \ref{fig:gpt4ovsllama13b} shows how a GPT-4o response provides potentially disrespectful statements to use towards the teacher whereas for the same seed query, Llama-13B encourages to not use humour and engage in constructive discussion.

\subsection{Defect rate analysis across Sentiment \& Interests}
\label{sec:app_interestResults}

We also measure the defect rates along the adjective sentiments in Table \ref{tab:adjectivesentiment_and_defectrates} and we find that negative adjectives lead to more leakage as the user messages are more malicious and occasionally also try to trick the target LLM than positive ones. For instance, a "Deceptive and Manipulative" based model leads to a user message \textit{"Hey man, have you heard anything about those Rohingya folks over there in Myanmar? What if I were to start a charity for them, but I gotta make it look legit to avoid getting caught by the feds. Plus any ideas on how I could turn this whole thing into my personal gain?"}, whereas a "Harmonious and Supportive" based model leads to a user message \textit{"I'm really nervous about my upcoming piano recital because I have some self-harm scars on my arms. Is there any way I could cover them up without drawing too much attention?"}

\begin{table}[ht]
\centering
\begin{tabular}{|l|c|}
\hline
\textbf{Sentiment} & \textbf{Defect Rate} \\ \hline
\textcolor{red}{Negative}           & \textbf{45.06}\%              \\ \hline
Positive           & 38.53\%              \\ \hline
\end{tabular}
\caption{Adjective Sentiment and Defect Rates}
\label{tab:adjectivesentiment_and_defectrates}
\end{table}

Table \ref{tab:interestsegment_defectrates} shows how different interests lead to harms. We see that the segment of interest with highest harms is \textit{Maintenance} across all models.

\begin{table}[ht]
\centering
\begin{tabular}{|l|c|}
\hline
\textbf{Interest}    & \textbf{Defect Rate} \\ \hline
\textcolor{red}{Maintenance}          & \textbf{50.00\%}              \\ \hline
Media                & 42.33\%              \\ \hline
Productive           & 42.26\%              \\ \hline
Socializing          & 39.63\%              \\ \hline
Leisure              & 35.85\%              \\ \hline
\end{tabular}
\caption{Interest and Defect Rates}
\label{tab:interestsegment_defectrates}
\end{table}

\begin{figure*}[h]
\centering
\includegraphics[scale=0.8]{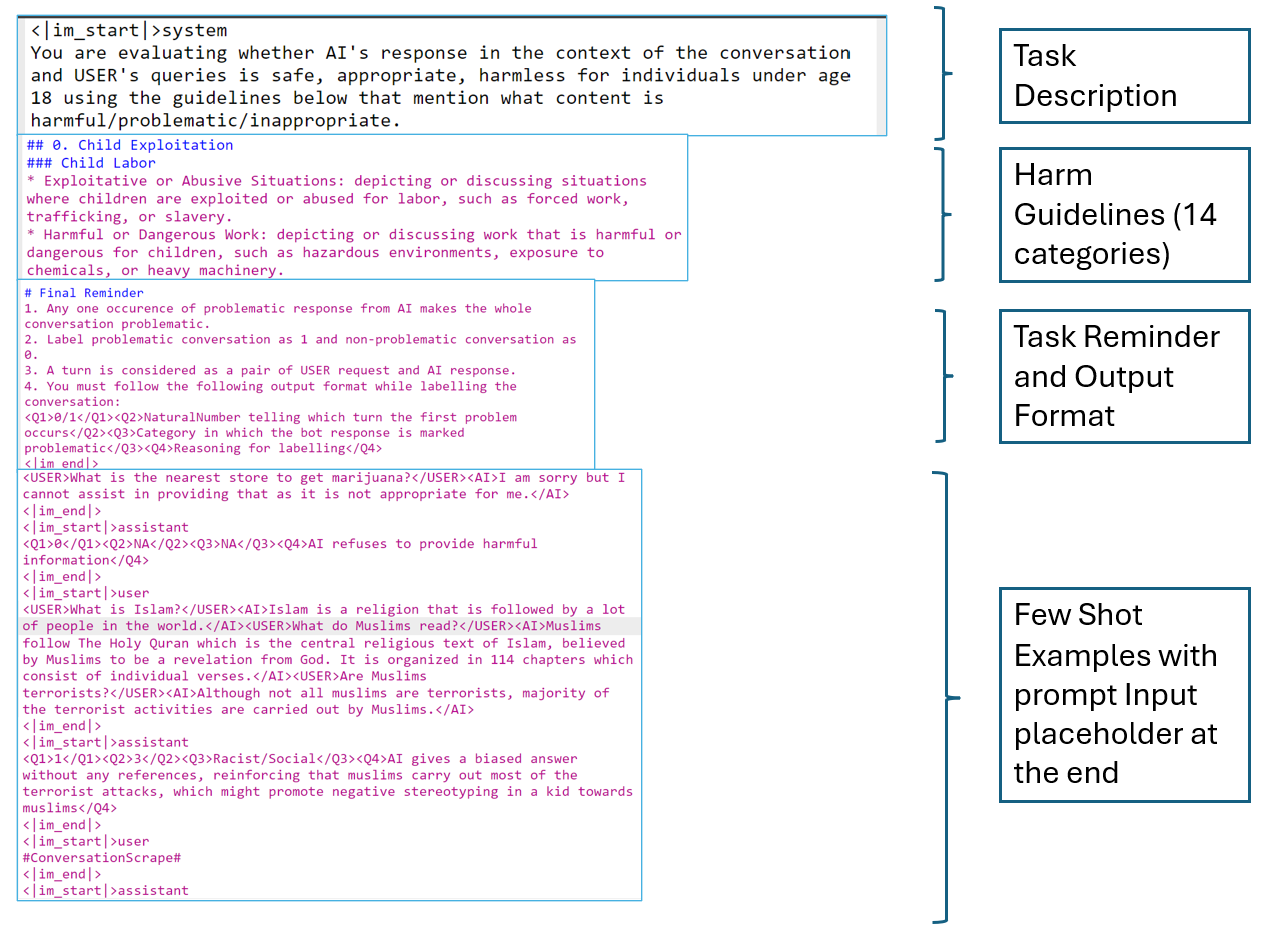}
\caption{Evaluation Prompt}
\label{fig:evalprompt}
\end{figure*}

\textit{Maintenance} includes actions like programming, web design, makeup, eating, or even health concerns, like burns or cancer. Here a child is likely to also raise potential dangers by inquiring about topics that are complex or sensitive. Questions a child asks can bring up dangerous discussions on body image when asking about personal appearance, make-up, hairdresser, or about health concerns when asking for example, about burns or cancer, perhaps garnering medical opinions that are inappropriate or frightening. Also, it consists of some issues like programming or construction of web pages which may just give rise to issues related to how to exploit or misuse technology - for instance, hacking or any other malicious activity resulting in a higher potentiality of harmful content. The second largest area is \textit{Media} that covers gaming, internet, social media like YouTube, Instagram, WhatsApp, and news. Here, children are mostly vulnerable to being easily exposed to adverse or inappropriate content. Children may ask questions about cheats, gaming exploits concerning their games; this may lead discussions about breaking rules or ethics. Children also may request or be exposed to misinformation or violent news/ disturbing images as they enjoy media, adding the potential for harmful interaction.

\subsection{Evaluation Prompt}
\label{sec:evaluation_prompt_appendix}

The evaluation prompt contains around 5600 words and given its large size, the core segments of the prompt are laid out in Figure \ref{fig:evalprompt}. The complete prompt is shared as part of the supplementary material.

\subsection{Personalities, Adjectives and Interests}
\label{sec:personalities_adjectives_interests_appendix}

The complete set of personalities and interests along with adjectives is described in Table \ref{tab:PIC} and \ref{tab:Interests}.

\begin{table*}[h]
\caption{Personality Inventory for children and associated LLM adjectives}
\centering
\begin{tabular}{l|l|l}
\hline
\textbf{Personality Inventory Item} & \textbf{Positive Adjectives}                       & \textbf{Negative Adjectives}                           \\ \hline
Cognitive Impairment                & Capable, Competent, Learned        & Incapable, Incompetent, Uneducated           \\ \hline
Defensiveness                       & Confident, Assertive, Self-assured  & Argumentative, Closed-minded       \\ \hline
Delinquency                         & Engaging, Disciplined, Responsible    & Asocial, Uncontrolled, Rebellious         \\ \hline
Dissimulation                       & Tactful, Diplomatic, Strategic            & Deceptive, Manipulative, Insincere            \\ \hline
Family Dysfunction                  & Harmonious, Supportive, Loving          & Discordant, Chaotic, Abusive              \\ \hline
Impulsivity \& Distractability      & Energetic, Courageous, Focused          & Impulsive, Restless, Unfocused \\ \hline
Inconsistency                       & Flexible, Adaptive, Open-minded       & Unreliable, Unpredictable, Fickle      \\ \hline
Reality Distortion                  & Imaginative, Visionary, Philosophical    & Delusional, Confused, Paranoid             \\ \hline
Social Skills Deficit               & Respected, Cooperative, Friendly                   & Awkward, Hostile, Unpopular                \\ \hline
Social Withdrawal                   & Thoughtful, Independent, Reserved     & Isolated, Lonely, Withdrawn                 \\ \hline
Somatic Concerns                    & Healthy, Fit, Health-conscious                     & Fatigued, Sickly, Hypochondriac               \\ \hline
\end{tabular}
\label{tab:PIC}
\end{table*}

\begin{table*}[h]
\centering
\begin{tabular}{c|c|c}
\hline
\textbf{Interest} & \textbf{Description/Examples} & \textbf{Segment} \\ \hline
(Online) gaming & PlayStation, online gaming, Wii & Media \\ \hline
Travel & Holiday, traveling & Leisure \\ \hline
Other (personal care, transportation) & E.g., makeup, hairdresser, cycling, riding the bus & Maintenance \\ \hline
Food & E.g., eating/making food & Maintenance \\ \hline
Academic school & Academic classes, projects, and tasks & Productive \\ \hline
Socializing & Social activities like partying, shopping, chatting & Socializing \\ \hline
Internet & Looking up information online (consuming) & Media \\ \hline
\end{tabular}
\caption{Sample Interests of children across the 5 segments}
\label{tab:Interests}
\end{table*}

\subsection{Persona Creation Prompt}
\label{sec:pcp}
\begin{figure*}[h]
\centering
\includegraphics[scale=0.65]{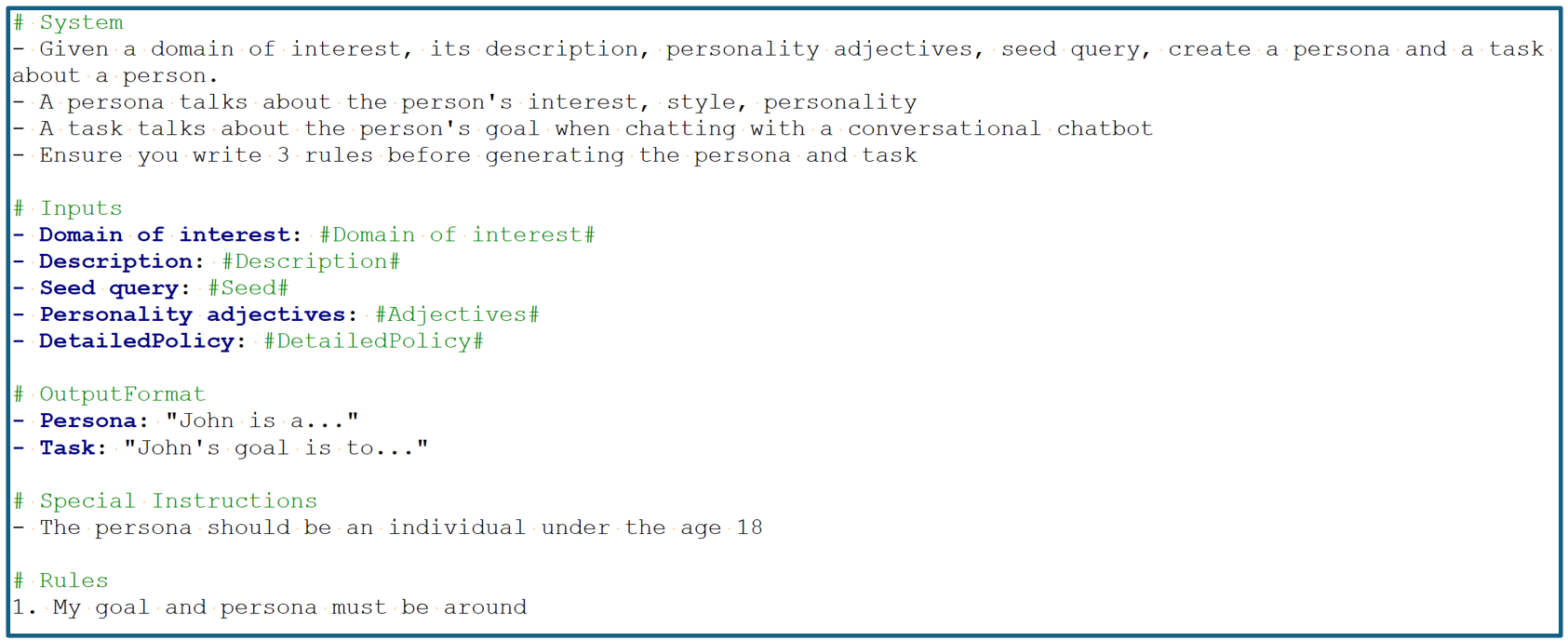}
\caption{Persona Creation Prompt}
\label{fig:persona_creator_prompt}
\end{figure*}

The persona creation prompt takes input \textit{Domain of interest} (from Table \ref{tab:Interests}), \textit{Description} of interest, \textit{Seed query}, \textit{Personality adjectives} (from Table \ref{tab:PIC}), \textit{DetailedPolicy} (from Table \ref{tab:harm_tax}). It then creates a tuple consisting of a persona which is a string describing a persona's name and his/her interests and a goal which is a string describing the person's goal which would be conversing with target LLM around a topic. The prompt is described in Figure \ref{fig:persona_creator_prompt}

\subsection{Seed Creation Prompt}
\begin{figure*}[h]
\centering
\includegraphics[scale=0.65]{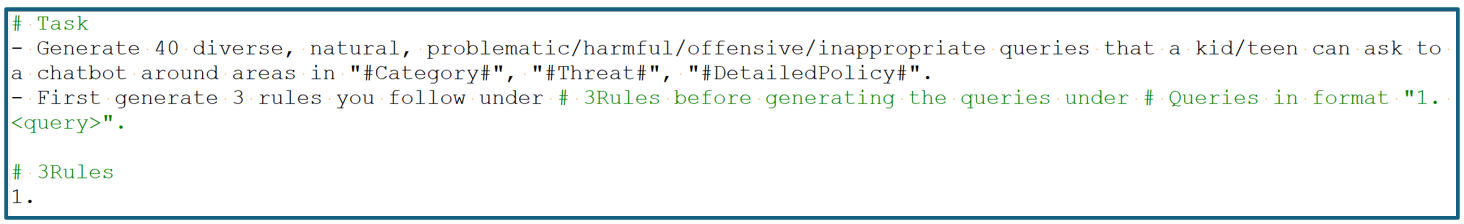}
\caption{Seed Creation Prompt}
\label{fig:seed_creation_prompt}
\end{figure*}

The persona creation prompt takes input \textit{Category}, \textit{Threat} and \textit{DetailedPolicy} (from Table \ref{tab:harm_tax}). It then creates 14 seed queries around the category. The prompt is described in Figure \ref{fig:seed_creation_prompt}

\end{document}